\documentclass[final, runningheads]{llncs}
\usepackage[T1]{fontenc}
\usepackage{graphicx}
\usepackage[inline]{enumitem}
\usepackage{hyperref}
\usepackage{float}

\usepackage{color}

\usepackage{amsmath}
\newcommand{\itadata}{\footnotesize \textsl{ITADATA2024: The 3$^{\text{rd}}$ Italian Conference on Big Data and Data Science}}
\usepackage{fancyhdr}
\fancypagestyle{empty}{\fancyhf{}\fancyhead[C]{\itadata}
  \fancyhead[R]{\includegraphics[width=.05\textwidth]{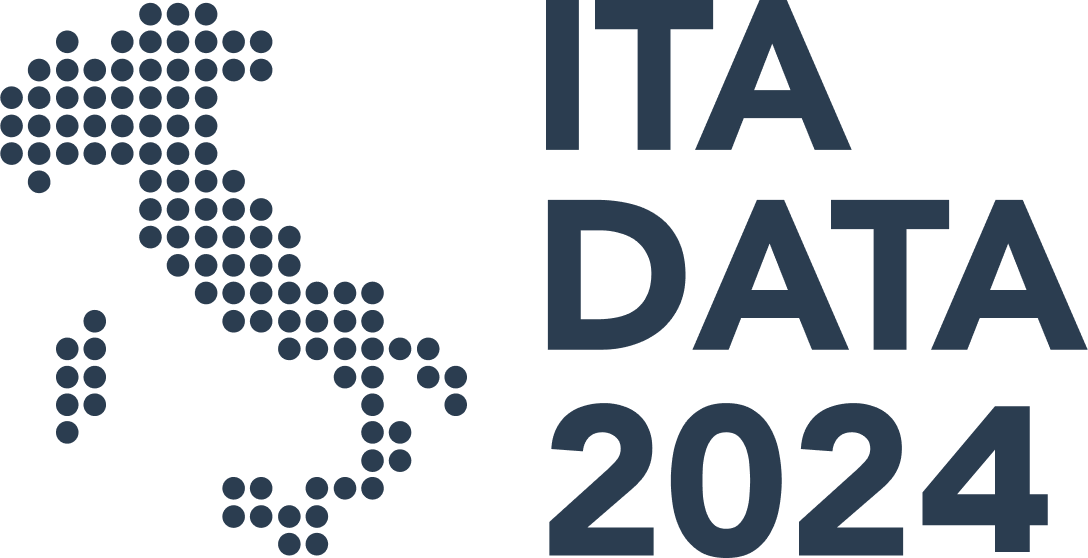}}}

\usepackage{wrapfig}
\usepackage{booktabs}
\usepackage{tabularx}
\usepackage{graphicx}
\usepackage{makecell}
\usepackage{array}
\usepackage[htt]{hyphenat}
\usepackage{stmaryrd}
\usepackage{multirow}

\makeatletter
\begin{document}
\title{From Text to Talent: A Pipeline for Extracting Insights from Candidate Profiles}

\titlerunning{From Text to Talent}
\author{Paolo Frazzetto\inst{1, 2}\orcidID{0000-0002-3227-0019} \and
Muhammad Uzair Ul Haq\inst{1, 2}\orcidID{0000-0001-9660-8982} \and Flavia Fabris\inst{1} \and
Alessandro Sperduti\inst{2}\orcidID{0000-0002-8686-850X}}
\authorrunning{P. Frazzetto et al.}
\institute{Amajor SB S.p.A, Noventa Padovana (PD), Italy\\
\email{\{rd1,flavia.fabris\}@amajorsb.com}\\ \and
Department of Mathematics
``Tullio Levi-Civita'', University of Padova, Padua, Italy\\
\email{\{paolo.frazzetto@phd.,muhammaduzair.ulhaq@phd., alessandro.sperduti@\}unipd.it}}
\maketitle              \begin{abstract}
The recruitment process is undergoing a significant transformation with the increasing use of machine learning and natural language processing techniques. While previous studies have focused on automating candidate selection, the role of multiple vacancies in this process remains understudied. This paper addresses this gap by proposing a novel pipeline that leverages Large Language Models and graph similarity measures to suggest ideal candidates for specific job openings. Our approach represents candidate profiles as multimodal embeddings, enabling the capture of nuanced relationships between job requirements and candidate attributes. The proposed approach has significant implications for the recruitment industry, enabling companies to streamline their hiring processes and identify top talent more efficiently. Our work contributes to the growing body of research on the application of machine learning in human resources, highlighting the potential of LLMs and graph-based methods in revolutionizing the recruitment landscape.

\end{abstract}

\section{Introduction}

Human Resource Management (HRM) is a strategic approach to managing an organization's most valuable asset---its people. It encompasses the policies, practices, and systems that influence employees' behavior, attitudes, and performance \cite{PAN2023100924}. The Human Resources (HR) department is responsible for implementing these strategies and handling functions such as recruitment, training, compensation, and employee relations \cite{strohmeier2022handbook}. HRM plays a crucial role in organizational success, as it directly impacts the recruitment of top talent, which in turn drives long-term growth, productivity, and competitiveness \cite{noe2016fundamentals}.
Traditionally, the hiring process relied heavily on manual assessment, including thorough Curriculum Vitae (CV) analysis, assessment questionnaires \cite{bailey2017hr}, and in-person interviews. However, the advent of digital technology has significantly transformed recruitment practices. Online job portals have revolutionized how organizations connect with potential candidates: these platforms offer substantial advantages to HR recruiters, allowing them to reach a vast talent pool efficiently. The widespread adoption of these online platforms has not only expanded the reach of job postings but has also paved the way for more sophisticated, data-driven recruitment strategies \cite{Marler}.

Although the proliferation of online recruitment platforms has significantly streamlined the job application process for candidates, this efficiency has inadvertently created new challenges for HR departments. The ease of application has led to a substantial increase in the volume of applications received for each job posting, resulting in an additional workload for recruiters. The task of shortlisting candidates has become increasingly time-consuming and labor-intensive, necessitating the development of automated systems to assist in this process.
In response to this need, there has been a growing interest in the application of various automated techniques to address the CV-job description (JD) matching problem, including natural language processing, machine learning (ML), and information retrieval methodologies \cite{FERNANDEZREYES201973,Yi2007MatchingRA}. However, the development and implementation of such systems face significant hurdles. One primary challenge is the scarcity of annotated datasets, which are crucial for training ML models. Due to privacy concerns and ethical considerations, many organizations are reluctant to release their recruitment data to the public domain, impeding research progress in this field.
Besides, the complex and non-standardized structure of CVs makes it difficult to extract relevant information consistently, often resulting in suboptimal matching with the job requirements. Additionally, while some researchers have employed rule-based systems for CV classification and clustering, due to their relative ease of development and reliance on keywords, these systems often lack the sophistication required to handle the nuances of complex documents or varied CV templates.
Furthermore, the limitations of keyword-based matching become apparent when dealing with the semantic richness of human language. Synonyms, context-dependent meanings, and industry-specific jargon can all lead to mismatches between equally qualified candidates and suitable job positions. This underscores the need for more advanced, context-aware systems that can understand the nuances of both job requirements and candidate qualifications. 

In this paper, we propose an innovative pipeline that harnesses the power of Large Language Models (LLMs) and Graph Neural Networks (GNNs) to gain insights into the recruitment process.
The pipeline consists of several key steps:
\begin{enumerate}
    \item \emph{Information Extraction:} We utilize LLMs to parse and extract multifaceted information from CVs, addressing the challenges of unstructured data. This includes soft and hard skills, education, work experience, and other relevant attributes.
    \item \emph{Embedding Generation:} The extracted information is then transformed into high-dimensional embeddings using the LLM. These embeddings capture the semantics of each attribute, allowing for more sophisticated comparisons.
    \item \emph{Heterogeneous Graph Construction:} Based on the similarity between these embeddings, we construct a heterogeneous graph. This graph represents the complex relationships among candidates.
    \item \emph{GNN Training and Inference:} We train different GNN architectures on this heterogeneous graph, enabling the model to learn and leverage the intricate patterns and relationships within the recruitment ecosystem for multiple selection processes. 
    \item \emph{Candidate Insights:} The task is to predict candidates' recruitment status for all the vacancies they applied to, framing this problem as multi-task and multi-class classification.
\end{enumerate}

To validate our approach, we test this pipeline on a proprietary, anonymized dataset comprising multiple job postings and candidate profiles. This real-world data allows us to demonstrate the effectiveness of our method, which opens more research opportunities. This approach has the potential to significantly streamline the hiring process, enabling recruiters with an additional tool to identify top talent more efficiently and effectively.

\section{Related Work}

Several studies are targeted at CV screening, but few are also related to multiple selections and their stages. Researchers have proposed automated systems with different approaches, which we present in this section.

Tian et al. \cite{Tian} proposed a resume classification system using Latent Semantic Analysis (LSA), Bidirectional Encoder Representations from Transformers (BERT), and Support Vector Machine (SVM). Similarly, Roy et al. \cite{ROY20202318} used Term Frequency-Inverse Document Frequency (TF-IDF) for feature extraction and cosine similarity for matching candidates with job postings.

Wang et al. \cite{Wang2021} developed a method to rank candidates using competence keywords, relying on TF-IDF and a Knowledge Graph (KG) built from a structured Competence Map (CMAP). Martinez et al. \cite{Martinez-Gil2020} proposed a framework that calculates the cost of each item in the CV relative to a taxonomy, ranking CVs based on their transformation cost with respect to the JD.

Kavas et al. \cite{Kavas} aligned CVs and job descriptions based on the European Skills, Competences, Qualifications and Occupations (ESCO) taxonomy. Phan et al. \cite{Phan} used the Computer Science Ontology (CSO) to classify documents and leveraged the Job-Onto ontology for CV-JD matching within the IT domain.

Wang et al. \cite{WangX} utilized BERT to encode textual information in resumes and vacancy requirements, combining it with graph embeddings and an attention interaction layer. Vanetik et al. \cite{Vanetik} extracted neural sentence representations, keywords, and named entities using BERT from resumes and vacancies, ranking based on vector representation distances.

Pessach et al. \cite{Pessach} introduced a comprehensive analytics framework for HR recruiters, combining a local prediction scheme using Variable-Order Bayesian Network (VOBN) and a global recruitment optimization model.

Jain et al. \cite{Jain} introduced CV and JD matching using topic modeling approaches such as LSA and Latent Dirichlet Allocation (LDA), combined with TextRank for CV summarization.

Pudasaini et al. \cite{Pudasaini} and Bothmer et al. \cite{Bothmer} employed word embedding techniques. \cite{Pudasaini} used word2vec with the CBOW model, while \cite{Bothmer} introduced Skill Scanner, which uses Sentence-BERT for skill vectorization and K-means clustering for grouping.

These approaches share several common themes. Many rely on text representation techniques, ranging from traditional methods like TF-IDF to more advanced approaches using BERT and other embedding models. There is a consistent focus on extracting meaningful features from unstructured text data in CVs and JDs. Several studies incorporate domain-specific knowledge, such as ontologies or taxonomies, to enhance matching accuracy. The use of similarity measures, particularly cosine similarity, is prevalent across multiple approaches. Ultimately, there is a trend towards more sophisticated machine learning techniques to capture complex relationships in HR data. 

\section{Methodology}
Our approach comprises four main stages. First, we detail our data collection process and the necessary preprocessing steps to prepare the raw data. Next, we explain how we leverage LLMs to generate contextual embeddings from the preprocessed data. The third stage involves the construction of a heterogeneous graph based on these embeddings and the computation of similarity measures. Finally, we describe the GNN architectures developed to learn this setting.

\subsection{Data Collection}
Due to privacy and ethical concerns, the access and the use of HR data in ML is a sensitive task \cite{AIAct,PAN2023100924}. Such a data collection process requires collaboration with organizations willing to share their recruitment data and provide access to their internal information systems.
This work has been made possible by the partnership and support of Amajor S.p.A SB, an Italian business development consulting firm. 
The dataset provided by Amajor has been collected over two years and includes $S=39$ completed selection processes of different clients, totaling $C=5461$ candidates. It comprises diverse roles and positions in multiple sectors, featuring a heterogeneous population of employees from various backgrounds and geographic locations. Nevertheless, some candidates have applied to multiple similar vacancies (or had been reconsidered by the recruiters for other suited positions), yielding to $6624$ unique candidate-application combinations. The selection processes had $170_{\pm 120}$ applicants, ranging from a minimum of $31$ to a maximum of $648$.

This research prioritizes ethical data handling and privacy compliance, and we obtained informed consent from all participants. To protect their privacy, we removed all personally identifiable information from the dataset (such as names, emails, phone numbers, ...). We also eliminated potential biasing factors such as age and gender to ensure fairness in our analysis. These characteristics allow us to test and potentially evaluate our proposed methodology, which traditionally cannot be examined due to insufficient data on such specific and diverse population groups.

\subsection{Dataset Features}
The dataset has two main components for each candidate: their CV and an assessment questionnaire results. The CV provides rich, unstructured text data about the candidate's qualifications, experience, and skills. Additionally, we utilized 18 numerical traits derived from Amajor's proprietary assessment questionnaire \cite{peronato2022}. These traits offer a standardized, quantitative perspective on each candidate's habits and behaviors, complementing the qualitative information in their CV. Each questionnaire trait takes numerical values in the range $[-100,100]$ and has been standard normalized. Missing values have been imputed by their median. 

\subsubsection{Target Labels}
For each selection process or vacancy, candidates were assigned an ordinal label $y$ reflecting their progression through the recruitment pipeline. The labels $y$ were defined as follows: $\textit{Applied/Rejected} \rightarrow 0, \textit{Screened/Interviewed} \\ \rightarrow 1, \textit{Offer Proposal} \rightarrow 2, \textit{Hired} \rightarrow 3$. 
This labeling scheme was developed after discussions with experienced recruiters. It was determined that predicting the higher categories (1, 2, and 3) would provide the most value for the recruitment process, as these represent candidates who progressed in the application stage.
\begin{wraptable}[8]{r}{0.4\textwidth}
\centering
\vspace{-10mm}
\caption{\small
Candidate labels (\%).}\label{tab:outcome_distribution}
\begin{tabularx}{0.35\textwidth}{>{\centering\arraybackslash}X>{\centering\arraybackslash}X>{\centering\arraybackslash}X>{\centering\arraybackslash}X>{\centering\arraybackslash}X}
\toprule
$y$ & Mean & Min & Max & Std \\
\midrule
0 & 94.77 & 78.87 & 99.22 & 3.88 \\
1 & 3.95 & 0.39 & 18.31 & 3.20 \\
2 & 1.47 & 0.27 & 4.17 & 1.09 \\
3 & 0.95 & 0.27 & 3.23 & 0.71 \\
\bottomrule
\end{tabularx}
\end{wraptable}
Our predictive modeling efforts thus focus on these more meaningful outcomes, aiming to identify candidates likely to reach the interview, offer, or hiring stages. Due to the nature of this setting, the classes are highly unbalanced, posing substantial challenges for model training. The class distribution statistics, grouped across all $39$ recruitment processes, are reported in Tab.~\ref{tab:outcome_distribution}.

\subsubsection{Feature Extraction}
\label{sec:feature_extraction}
Given that CVs are typically unstructured data, traditional rule-based approaches often struggle to parse and interpret the various formats accurately. Moreover, ML-based approaches rely on annotated datasets for information extraction \cite{uzair,Skillspan}. Therefore, to transform the unstructured text of CVs into structured data, we leveraged LLMs (namely, GPT-4 \cite{openai2023gpt4}) since their advanced natural language processing capabilities proved to be effective for information extraction \cite{Tang2024}. Also, the GPT-4 model can efficiently parse through diverse document formats, identifying relevant entities from the text \cite{wei2024chatie}. 

In this setting, we harnessed the capabilities of GPT-4 to extract five key entities from each CV: $\mathcal{E} = \{\textit{Soft Skills}, \textit{Hard Skills}, \textit{Industry Sector}, \textit{Education},\allowbreak \textit{Language Skills}\}$, thereby creating a structured representation of each candidate's profile containing the keywords related to these entities. We denote one of these entities with $\epsilon\in\mathcal{E}$. To achieve this, we first normalize the text by removing extra spaces, special characters, and sensitive private data. Then, we use GPT-4 \cite{openai2023gpt4} to extract entities using the prompting approach and store them in a structured data format for all CVs.

\subsection{Embeddings Generation}

To create vector representations of all features, we employed OpenAI's \texttt{text-embedding-3-large} model \cite{OpenAI2024textembedding3}, which represents the state-of-the-art in embedding technology. This model generates a vector $\mathbf{v}_{\epsilon,i}\in\mathrm{R}^{768}$ for each textual feature belonging to category $\epsilon$ for candidate $i$, allowing us to capture rich semantic information.
In this real-world scenario, the number of entities varies across candidates. For instance, Candidate A might possess five soft skills, while Candidate B has only three. To accommodate this variability, we represent each CV for candidate $i$ ($\text{CV}_i$) as a \emph{set of sets} of the five entity categories $\mathcal{E}$, where each entity category is encoded as the sets of its vector embeddings $E_{\epsilon,i} = \{\mathbf{v}_{\epsilon, i}\}$. In mathematical terms, $\text{CV}_i = \{E_{\epsilon,i}\; | \; \epsilon \in \mathcal{E}\}.$ It is worth mentioning that the vector embeddings allow us to overcome the issues of synonyms and CVs written in different languages since such words are closer in the embedding space.

\subsection{Graph Construction and Similarity Measures}
Our approach is predicated on the assumption that candidates with comparable CVs are likely to be similarly suited for the same vacancies. To operationalize this concept, we aimed to design a robust similarity measure, ultimately constructing a graph of candidates \cite{Frazzetto2023EnhancingHR}. Given that each CV is represented by a set of embeddings for all entities with varying cardinalities, we employed an approximate nearest neighbor (ANN) algorithm \cite{douze2024faiss} to identify the k-nearest neighbors (with $k=10$) for each entity $\epsilon$ across all CVs. See Tab.~\ref{tab:nearest_neighbors} for an example of retrieved neighbors, showcasing the efficacy of this method. \\
\newcolumntype{Y}{>{\centering\arraybackslash}X}
\newcolumntype{M}[1]{>{\centering\arraybackslash}m{#1}}
\begin{table}[htbp]
\centering
\small
\setlength{\abovecaptionskip}{-18pt}
\caption{Example of Top 5-NN based on a given query for all entities.}
\label{tab:nearest_neighbors}
\begin{tabularx}{\textwidth}{@{}lX@{}}
\toprule
\textbf{Entity} & \textbf{Nearest Neighbors} \\
\midrule
\makecell[l]{Soft Skills: \\ \textit{ communication}} & \makecell[l]{communications, comunication, communications and \\relations, communicating, communication and writing} \\ 
\makecell[l]{Hard Skills: \\ \textit{ python}} & \makecell[l]{python programming, python), coding (python, \\programming languages (python, python (numpy} \\ 
\makecell[l]{Industry Sector: \\ \textit{ startups}} & \makecell[lc]{start ups, start up companies, technology startups,\\innovative start ups, technology startup} \\ 
\makecell[l]{Education: \\ \textit{ management diploma}} & \makecell[lc]{master in management, marketing management diploma,\\master degree in management, master in general\\management, professional diploma in management} \\ 
\makecell[l]{Languages: \\ \textit{ english}} & \makecell[lc]{english (good), english (medium), english (school), \\english (native), english language} \\ 

\bottomrule
\end{tabularx}

\end{table} 
\vspace{-3ex}
Formally, we define the kNN vectors retrieved for one embedding $\mathbf{v}_{\epsilon,i}$ as $ \operatorname{kNN}(\mathbf{v}_{\epsilon,i})$.
Inspired by the Jaccard Index $J$ for sets \cite{wang2024jaccard}, we compute the overlap between two candidates $i$ and $j$ for a given entity $\epsilon$ as
\begin{equation}
    J_\epsilon(i,j) = \frac{\sum_{\mathbf{v}_{\epsilon,i}, \mathbf{v}_{\epsilon,j}}\llbracket \operatorname{kNN}(\mathbf{v}_{\epsilon,i})\subseteq\operatorname{kNN}(\mathbf{v}_{\epsilon,j})\rrbracket}{|E_{\epsilon, i}| + |E_{\epsilon, j}|},
\end{equation}
where $\llbracket \cdot \rrbracket$ is the indicator function. $J_\epsilon(i,j)$ is the count of all embeddings that share at least one neighbor, normalized by the total amount of the embeddings. The similarity is then computed as:
\begin{equation}
    \label{eq:similarity}
    \operatorname{sim}_\epsilon(i,j) = \operatorname{max}(1 - e^{(-\lambda J_\epsilon(i,j))} - \theta, 0)
\end{equation}
\begin{wraptable}[8]{r}{0.4\textwidth}
    \centering
\vspace{-4mm}
\caption{\small Amount of edges.}
\label{tab:edge_counts}
\begin{tabularx}{0.35\textwidth}{Xr}
\toprule
\textbf{Category} & \textbf{\#Edges} \\
\midrule
Soft Skills & 1,247,845 \\
Hard Skills & 162,605 \\
Industry Sector & 478,838 \\
Education & 287,724 \\
Languages & 7,723,929 \\
\bottomrule
\end{tabularx}

\end{wraptable}
so that $\operatorname{sim}_\epsilon(i,j) \in [0, 1-\theta]$, $\lambda$ is a scaling factor and $\theta$ is a tunable threshold to discard smaller values. 
Running this algorithm for all entities, we obtained a \emph{heterogeneous graph} $G$ of candidates connected by different types of weighted edges according to their similarity.
We set $\lambda=2$ and $\theta=0.2$, obtaining a graph with $9,900,941$ edges, grouped as in Tab.~\ref{tab:edge_counts}.
\subsection{Model Architecture}
In our graph-based approach, we employed Heterogeneous Graph Convolution as the core component of our model architecture \cite{HeteroConv}. Within this framework, we implemented two popular graph convolution methods: Graph Convolutional Networks (GCN) \cite{kipf2017GCN} and Relational Graph Convolutional Networks (RGCN) \cite{RGCN}. GCN provides a powerful mechanism for aggregating information from local graph neighborhoods. In this heterogeneous formulation, each edge type is treated individually, so the message passing happens on all five induced subgraphs. Finally, each hidden representation for each type is summed. RGCN extends this capability by explicitly modeling different types of edges in the graph, but contrary to GCN, the original implementation does not consider edge weights. \subsubsection{Learning Framework}
The final goal would be to predict the HR labels $y$ so that new vacancies and candidates can be added to the graph in a future stage. 
We treated these selections in a \emph{multi-task} learning approach to exploit their shared structure and learn relationships among candidates and vacancies. Another viable solution would be to train a classifier by first merging all selections \cite{Frazzetto2023EnhancingHR}, or train a model on the subgraphs induced by each selection. These options are left open to future works. Additionally, given the nature of our recruitment outcome data, which consists of four ordered classes, we explored two distinct problem formulations. First, we approached the task as  \emph{ordinal regression}, recognizing the inherent order and progressive nature of the recruitment stages. Alternatively, we framed the problem as a \emph{multi-label classification} task, where each candidate could be assigned one or more labels corresponding to the stages they reached. This approach offers flexibility when a candidate might skip certain stages, or the recruitment process does not strictly follow a linear progression. 

\section{Experimental Setup}
The graph $G$ entails all $C=5461$ candidates, whose features are the $18$ numerical questionnaire's traits; its edges are given by their CVs similarity according to the five entities embeddings, each one weighted by its similarity score. Every candidate has a categorical label belonging to at least one of the $S=39$ job selections. Dealing with a single graph, we fall into the \emph{transductive} learning case. We split the nodes evenly for all tasks with $80\%/20\%$ splits for train/test sets. 
\begin{wraptable}[10]{r}{0.5\textwidth}
    \centering
    \vspace{-10mm}
    \caption{Hyperparameters grid.}
    \label{tab:optuna_hyperparameters}
    \begin{tabularx}{0.5\textwidth}{lX}
        \toprule
        \textbf{Hyperpar.} & \textbf{Range/Values} \\
        \midrule
        Hidden Units & 16 to 64  \\
        \#Deep Layers & 1 to 5 \\
        Learning Rate & $10^{-4}$ to $10^{-1}$  \\
        Activation Function & \makecell[l]{LeakyRelu, Elu,\\ Tanh, Sigmoid} \\
        \bottomrule
    \end{tabularx}
\end{wraptable}
Due to class imbalances, we made sure the splits were properly stratified. Rare labels $y=4$ were also randomly split with the same ratio.
The models have been implemented with the \texttt{PyTorch Geometric} library \cite{PyG}.
We ran each trial for 300 epochs with the Adam optimizer. The hyperparameters were chosen over the values in  Tab.~\ref{tab:optuna_hyperparameters} with the \texttt{Optuna} library \cite{optuna_2019} for 100 trials. 

\section{Results}
\begin{table}[h]
    \centering
    \small
    \setlength{\abovecaptionskip}{-8pt}
    \caption{Experiment Results for RGCN and GCN with Ordinal Regression and Multilabel Learning}\label{tab:experiment_results}
    \begin{tabularx}{0.85\textwidth}{ll*5{>{\centering\arraybackslash}X}}
        \toprule
        \textbf{Model} & \textbf{Learning} & \textbf{Acc.} & \textbf{MAE} & \textbf{RMSE} & \textbf{F1} & \textbf{AUC} \\
        \midrule
        \multirow{2}*{\textbf{RGCN}} & Ordinal & 25.2 & 0.532 & 0.924 & 0.729 & 0.566 \\
         & Multilabel & 20.1 & 1.07 & 1.58 & 0.615 & 0.503 \\ \midrule
        \multirow{2}*{\textbf{GCN}} & Ordinal & 27.4 & 0.565 & 0.900 & 0.684 & 0.539 \\
         & Multilabel & 30.1 & 0.662 & 1.29 & 0.796 & 0.606 \\
        \bottomrule
    \end{tabularx}
\end{table}
\noindent The results in Tab.~\ref{tab:experiment_results} demonstrate the performance of the RGCN and GCN models under two different learning scenarios: ordinal regression and multi-label learning. Due to classes' distribution, we report the \emph{balanced} accuracy and \emph{weighted} F1-score. We also grouped classes $(0,1)$ and $(2,3)$ to compute the AUC, as for a binary classification setting.

The GCN model generally outperforms the RGCN model in both learning types. Specifically, for the ordinal regression task, the GCN model achieved a higher accuracy compared to the RGCN. The GCN model also demonstrated a lower RMSE and a competitive F1 score. Despite these gains, the RGCN model exhibited a slightly better binary AUC, indicating a marginally better binary classification capability.

In the multi-label learning scenario, the GCN model again showed superior performance in terms of accuracy, which is significantly higher than the RGCN. The GCN's F1 weighted score and AUC  highlight its robustness in handling multiple labels, which is crucial for such tasks requiring the prediction of several categories simultaneously.

While these results are not overwhelming in absolute terms, they are nonetheless promising. In particular, when considering the challenges posed by real-world, noisy, and imbalanced data in this complex learning setting. These findings not only demonstrate the feasibility of the proposed approach but also open up numerous avenues for future research directions, suggesting that further refinements could yield significant improvements for recruitment decision support systems.

\section{Conclusion}
This study explored the integration of GNNs and LLMs to support recruiters in personnel selections. By leveraging the strengths of GNNs in capturing relational data and the advanced text-understanding capabilities of LLMs, we developed a pipeline to process this type of HR data. The experimental results demonstrated that our approach can be applied in this real-world scenario. Finally, it showcases the potential of combining graph-based and language-based models for complex classification tasks, offering a robust solution for CV analysis.

\subsubsection{Future Research Direction}
The primary objective of this study is to establish a robust pipeline capable of handling the complexities of real-world HR data. While this paper explores and evaluates several architectural choices, learning strategies, and models, it is acknowledged that the vast landscape of possible approaches leaves ample room for future research and improvements. 
As a next step, we plan to add more candidates and selections and investigate the learned embeddings and their relations, as well as the graph structure. It is straightforward to process Job Descriptions in a similar fashion and add them as different node types, as with CV-JD matching. This would also ease the burden of multi-task learning, translating the problem into heterogeneous node ordinal regression. It is finally worth investigating other types of heterogeneous GNNs, and solutions for unbalanced target distributions.

\bibliographystyle{splncs04}
\bibliography{bibliography}

\begin{thebibliography}{10}
\providecommand{\url}[1]{\texttt{#1}}
\providecommand{\urlprefix}{URL }
\providecommand{\doi}[1]{https://doi.org/#1}

\bibitem{AIAct}
Regulation (eu) 2024 of the european parliament and of the council of laying
  down harmonised rules on artificial intelligence and amending regulations
  (ec) no 300/2008, (eu) no 167/2013, (eu) no 168/2013, (eu) 2018/858, (eu)
  2018/1139 and (eu) 2019/2144 and directives 2014/90/eu, (eu) 2016/797 and
  (eu) 2020/1828 (artificial intelligence act). European Commission (2024),
  \url{https://artificialintelligenceact.eu/the-act/}

\bibitem{openai2023gpt4}
Achiam, J., Adler, S., Agarwal, S., Ahmad, L., Akkaya, I., Aleman, F.L.,
  Almeida, D., Altenschmidt, J., Altman, S., Anadkat, S., et~al.: Gpt-4
  technical report. arXiv preprint arXiv:2303.08774  (2023)

\bibitem{optuna_2019}
Akiba, T., Sano, S., Yanase, T., Ohta, T., Koyama, M.: Optuna: A
  next-generation hyperparameter optimization framework. In: Proceedings of the
  25th {ACM} {SIGKDD} International Conference on Knowledge Discovery and Data
  Mining (2019)

\bibitem{bailey2017hr}
Bailey, R.: Hr applications of psychometrics. Psychometric Testing: Critical
  Perspectives pp. 85--111 (2017)

\bibitem{Bothmer}
Bothmer, K., Schlippe, T.: Skill scanner: Connecting and supporting employers,
  job seekers and educational institutions with an ai-based recommendation
  system. In: Guralnick, D., Auer, M.E., Poce, A. (eds.) Innovative Approaches
  to Technology-Enhanced Learning for the Workplace and Higher Education. pp.
  69--80. Springer International Publishing, Cham (2023)

\bibitem{douze2024faiss}
Douze, M., Guzhva, A., Deng, C., Johnson, J., Szilvasy, G., Mazaré, P.E.,
  Lomeli, M., Hosseini, L., Jégou, H.: The faiss library  (2024)

\bibitem{FERNANDEZREYES201973}
Fernández-Reyes, F.C., Shinde, S.: Cv retrieval system based on job
  description matching using hybrid word embeddings. Computer Speech \&
  Language  \textbf{56},  73--79 (2019).
  \doi{https://doi.org/10.1016/j.csl.2019.01.003},
  \url{https://www.sciencedirect.com/science/article/pii/S0885230817302851}

\bibitem{PyG}
Fey, M., Lenssen, J.E.: Fast graph representation learning with {PyTorch
  Geometric}. In: ICLR Workshop on Representation Learning on Graphs and
  Manifolds (2019)

\bibitem{Frazzetto2023EnhancingHR}
Frazzetto, P., Uzair-Ul-Haq, M., Sperduti, A.: Enhancing human resources
  through data science: a case in recruiting. In: itaDATA (2023),
  \url{https://api.semanticscholar.org/CorpusID:267035831}

\bibitem{Jain}
Jain, L., Harsha~Vardhan, M.A., Kathiresan, G., Narayan, A.: Optimizing people
  sourcing through semantic matching of job description documents and candidate
  profile using improved topic modelling techniques. In: Chiplunkar, N.N.,
  Fukao, T. (eds.) Advances in Artificial Intelligence and Data Engineering.
  pp. 899--908. Springer Nature Singapore, Singapore (2021)

\bibitem{Kavas}
Kavas, H., Serra-Vidal, M., Wanner, L.: Job offer and applicant cv
  classification using rich information from a labour market taxonomy (2023),
  \url{https://ssrn.com/abstract=4519766}, available at SSRN:
  https://ssrn.com/abstract=4519766 or http://dx.doi.org/10.2139/ssrn.4519766

\bibitem{kipf2017GCN}
Kipf, T.N., Welling, M.: Semi-supervised classification with graph
  convolutional networks (2017)

\bibitem{Marler}
Marler, J., Fisher, S.: An evidence-based review of e-hrm and strategic human
  re-source management. Human Resource Management Review  \textbf{23},  18–36
  (03 2013). \doi{10.1016/j.hrmr.2012.06.002}

\bibitem{Martinez-Gil2020}
Martinez-Gil, J., Paoletti, A.L., Pichler, M.: A novel approach for learning
  how to automatically match job offers and candidate profiles. Information
  Systems Frontiers  \textbf{22}(6),  1265--1274 (December 2020).
  \doi{10.1007/s10796-019-09929-7},
  \url{https://doi.org/10.1007/s10796-019-09929-7}

\bibitem{noe2016fundamentals}
Noe, R.A., Hollenbeck, J.R., Gerhart, B.A., Wright, P.M.: Fundamentals of human
  resource management. McGraw-Hill Education New York, NY (2016)

\bibitem{OpenAI2024textembedding3}
OpenAI: text-embedding-3-large model (2024),
  \url{https://www.openai.com/models/text-embedding-3-large}, accessed:
  2024-06-28

\bibitem{PAN2023100924}
Pan, Y., Froese, F.J.: An interdisciplinary review of ai and hrm: Challenges
  and future directions. Human Resource Management Review  \textbf{33}(1),
  100924 (2023). \doi{https://doi.org/10.1016/j.hrmr.2022.100924},
  \url{https://www.sciencedirect.com/science/article/pii/S1053482222000420}

\bibitem{peronato2022}
Peronato, E., Fabris, F., D'Agnolo, N., D'Orazio, R.: Entrepreneurial values as
  a key for csr in smes. Presented at the XXXIII ISPIM, Copenhagen, 2022
  (2022),
  \url{https://www.conferencesubmissions.com/ispim/copenhagen2022/index.html}

\bibitem{Pessach}
Pessach, D., Singer, G., Avrahami, D., {Chalutz Ben-Gal}, H., Shmueli, E.,
  Ben-Gal, I.: Employees recruitment: A prescriptive analytics approach via
  machine learning and mathematical programming. Decision Support Systems
  \textbf{134},  113290 (2020).
  \doi{https://doi.org/10.1016/j.dss.2020.113290},
  \url{https://www.sciencedirect.com/science/article/pii/S0167923620300452}

\bibitem{Phan}
Phan, T.T., Pham, V.Q., Nguyen, H.D., Huynh, A.T., Tran, D.A., Pham, V.T.:
  Ontology-based resume searching system for job applicants in information
  technology. In: Fujita, H., Selamat, A., Lin, J.C.W., Ali, M. (eds.) Advances
  and Trends in Artificial Intelligence. Artificial Intelligence Practices. pp.
  261--273. Springer International Publishing, Cham (2021)

\bibitem{Pudasaini}
Pudasaini, S., Shakya, S., Lamichhane, S., Adhikari, S., Tamang, A., Adhikari,
  S.: Scoring of Resume and Job Description Using Word2vec and Matching Them
  Using Gale–Shapley Algorithm, pp. 705--713. Springer International
  Publishing (01 2022). \doi{10.1007/978-981-16-2126-0_55}

\bibitem{ROY20202318}
Roy, P.K., Chowdhary, S.S., Bhatia, R.: A machine learning approach for
  automation of resume recommendation system. Procedia Computer Science
  \textbf{167},  2318--2327 (2020).
  \doi{https://doi.org/10.1016/j.procs.2020.03.284},
  \url{https://www.sciencedirect.com/science/article/pii/S187705092030750X},
  international Conference on Computational Intelligence and Data Science

\bibitem{RGCN}
Schlichtkrull, M., Kipf, T.N., Bloem, P., Van Den~Berg, R., Titov, I., Welling,
  M.: Modeling relational data with graph convolutional networks. In: The
  semantic web: 15th international conference, ESWC 2018, Heraklion, Crete,
  Greece, June 3--7, 2018, proceedings 15. pp. 593--607. Springer (2018)

\bibitem{strohmeier2022handbook}
Strohmeier, S.: Handbook of Research on Artificial Intelligence in Human
  Resource Management. Edward Elgar Publishing (2022)

\bibitem{Tang2024}
Tang, Y., Xiao, Z., Li, X., Zhang, Q., Chan, E.W., Wong, I.C.: Large language
  model in medical information extraction from titles and abstracts with prompt
  engineering strategies: A comparative study of gpt-3.5 and gpt-4. medRxiv
  (2024). \doi{10.1101/2024.03.20.24304572},
  \url{https://www.medrxiv.org/content/early/2024/03/21/2024.03.20.24304572}

\bibitem{Tian}
Tian, X., Pavur, R., Han, H., Zhang, L.: A machine learning-based human
  resources recruitment system for business process management: using lsa, bert
  and svm. Business Process Management Journal  \textbf{29}(1),  202--222 (2023
  2023),
  \url{https://www.proquest.com/scholarly-journals/machine-learning-based-human-resources/docview/2764035408/se-2},
  copyright - © Emerald Publishing Limited

\bibitem{uzair}
Ul~Haq, M.U., Frazzetto, P., Sperduti, A., Da~San~Martino, G.: Improving soft
  skill extraction via data augmentation and embedding manipulation. In:
  Proceedings of the 39th ACM/SIGAPP Symposium on Applied Computing. p.
  987–996. SAC '24, Association for Computing Machinery, New York, NY, USA
  (2024). \doi{10.1145/3605098.3636010},
  \url{https://doi.org/10.1145/3605098.3636010}

\bibitem{Vanetik}
Vanetik, N., Kogan, G.: Job vacancy ranking with sentence embeddings, keywords,
  and named entities. Information  \textbf{14}(8) (2023).
  \doi{10.3390/info14080468}, \url{https://www.mdpi.com/2078-2489/14/8/468}

\bibitem{HeteroConv}
Wang, X., Bo, D., Shi, C., Fan, S., Ye, Y., Philip, S.Y.: A survey on
  heterogeneous graph embedding: methods, techniques, applications and sources.
  IEEE Transactions on Big Data  \textbf{9}(2),  415--436 (2022)

\bibitem{WangX}
Wang, X., Jiang, Z., Peng, L.: A deep-learning-inspired person-job matching
  model based on sentence vectors and subject-term graphs. Complexity
  \textbf{2021}(1),  6206288 (2021).
  \doi{https://doi.org/10.1155/2021/6206288},
  \url{https://onlinelibrary.wiley.com/doi/abs/10.1155/2021/6206288}

\bibitem{Wang2021}
Wang, Y., Allouache, Y., Joubert, C.: Analysing {CV} corpus for finding
  suitable candidates using knowledge graph and {BERT}. In: DBKDA 2021, The
  Thirteenth International Conference on Advances in Databases, Knowledge, and
  Data Applications. Valencia, Spain (May 2021), hal-03325062

\bibitem{wang2024jaccard}
Wang, Z., Ning, X., Blaschko, M.: Jaccard metric losses: Optimizing the jaccard
  index with soft labels. Advances in Neural Information Processing Systems
  \textbf{36} (2024)

\bibitem{wei2024chatie}
Wei, X., Cui, X., Cheng, N., Wang, X., Zhang, X., Huang, S., Xie, P., Xu, J.,
  Chen, Y., Zhang, M., Jiang, Y., Han, W.: Chatie: Zero-shot information
  extraction via chatting with chatgpt (2024),
  \url{https://arxiv.org/abs/2302.10205}

\bibitem{Yi2007MatchingRA}
Yi, X., Allan, J., Croft, W.B.: Matching resumes and jobs based on relevance
  models. In: Annual International ACM SIGIR Conference on Research and
  Development in Information Retrieval (2007)

\bibitem{Skillspan}
Zhang, M., Jensen, K.N., Sonniks, S.D., Plank, B.: Skillspan: Hard and soft
  skill extraction from english job postings. In: North American Chapter of the
  Association for Computational Linguistics (2022)

\end{thebibliography}

\end{document}